\documentclass[12pt,preprint]{aastex}

\slugcomment{Submitted to xx}
\shortauthors{Brinkworth et al.}
\shorttitle{The Dusty Disk Around SDSS1228+1040}

\begin{document}

\title{
A Dusty Component to the Gaseous Debris Disk around the White Dwarf SDSS J1228+1040
}
 
\author{C. S. Brinkworth,\altaffilmark{1} 
B. T. G\"ansicke,\altaffilmark{2} T. R. Marsh,\altaffilmark{2} D. W. Hoard, \altaffilmark{1}
C. Tappert,\altaffilmark{3}
}
\altaffiltext{1}{Spitzer Science Center, California Institute of Technology,
Pasadena, CA 91125}
\altaffiltext{2}{Dept of Physics and Astronomy, University of Warwick, Warwick,CV4 7AL,  United Kingdom}
\altaffiltext{3}{Dpto de Astronom\'{\i}a y Astrof\'{\i}sica, Pontificia Universidad Cat\'olica de Chile, Casilla 306, Santiago 22, Chile}

\begin{abstract}
We present ISAAC spectroscopy and ISAAC, UKIDSS and \emph{Spitzer
  Space Telescope} broad-band photometry of SDSS J1228+1040 -- a white
dwarf for which evidence of a gaseous metal-rich circumstellar disk
has previously been found from optical emission lines.  The data show
a clear excess in the near- and mid-infrared, providing compelling
evidence for the presence of dust in addition to the previously
identified gaseous debris disk around the star. The infrared excess
can be modelled in terms of an optically thick but geometrically thin
disk. We find that the inner disk temperatures must be relatively high
($\sim$1700\,K) in order to fit the SED in the near-infrared. These data
provide the first evidence for the co-existence of both gas and dust
in a disk around a white dwarf, and show that their presence is
possible even around moderately hot ($\sim$22,000\,K) stars.

\end{abstract}


\section{Introduction}

Despite the numerous extrasolar planets discovered over the last
decade, the fate of these systems during the final stages of the
parent star's evolution is still uncertain. A clue to their destiny
was discovered during a search for cool companions to white
dwarfs. G29--38 was found to display a strong near-infrared excess
\citep{zuckerman87}, but deep imaging and asteroseismological studies
of the system ruled out the presence of a brown dwarf
companion. Mid-infrared photometry showed that this excess is even
more pronounced at 10 $\mu$m suggesting that the emission is due to
dust around the white dwarf \citep{graham90, telesco90,
  tokunaga90}. \emph{Spitzer Space Telescope} observations
spectroscopically confirmed the presence of circumstellar dust around
G29--38 \citep{reach05}, which has since been modelled as a flat,
opaque, circumstellar disk \citep{vonhippel07}. The atmosphere of
G29--38 has been found to be enriched with metals, implying that the
star is currently accreting \citep{koester97}, probably from this
dusty disk.

The more likely origin for the disk is the tidal disruption of either
comets or asteroids \citep{debes02, jura03}, with asteroids the more
probable candidates, as they can explain the large amount of metals
accreted by the white dwarfs. Comets are also dynamically less
favourable, given their eccentric orbits and long orbital periods. The
approach used by \citet{vonhippel07} to model the disk was based on a
simple, optically thick, geometrically thin dusty disk model described
in \citet{jura03}.

A further 11 white dwarfs have been found to display a similar
infrared excess \citep{becklin05, kilic05, kilic06, vonhippel07,
  jura07, jura08, farihi08a, farihi08b}, most of which are cool,
metal-rich DAZ white dwarfs, with effective temperatures
$T_\mathrm{eff} <$ 15000\,K (with the exceptions of PG\,1015+161,
which has $T_\mathrm{eff} \sim$19300\,K, and PG\,1457-086 with
$T_\mathrm{eff} \sim$ 20400\,K), leading \citet{kilic06} to suggest
that dusty disks around white dwarfs should be sublimated for
$T_\mathrm{eff} \geq$ 16000 -- 20000\,K. More recently,
\citet{gaensicke06, gaensicke07, gaensicke08} identified three more
moderately hot white dwarfs showing double-peaked calcium lines
indicative of a rotating circumstellar debris disk -
SDSS\,J122859.93+104032.9 (hereafter SDSS\,1228+1040),
SDSS\,J104341.53+085558.2 and SDSS\,J084539.17+225728.0.

Here we report the detection of a significant infrared excess in one
of these systems, SDSS\,1228+1040. The white dwarf was identified by
\cite{gaensicke06} in Data Release 6 of the Sloan Digital Sky Survey
\citep{adelman-mccarthy06}. \citet{gaensicke06} noted very unusual
emission lines of the Ca II 850-866\,nm triplet, as well as weaker
emission lines of Fe II at 502\,nm and 517\,nm. The line profiles of
the Ca II triplet display a double-peaked morphology, indicative of a
gaseous, rotating disk structure. Time-resolved spectroscopy ruled out
any significant radial velocity variations, and the absence of Balmer
and helium emission lines implied that the gaseous disk around the
white dwarf must be extremely deficient in volatile materials, ruling
out the presence of a stellar binary companion. The velocity of the Ca
II line peaks indicate that the outer radius of the gaseous
circumstellar disk is $\simeq1.2$\,R$_{\odot}$.

\section{Observations and Data Reduction}

\subsection{Spitzer Space Telescope data}
We were awarded 2.4 hours of \emph{Spitzer Space Telescope} time in
Cycle-4 (P40048), plus 3.0 hours in Cycle-5 (P50118) to search for a
dusty component to the gaseous circumstellar disk around
SDSS\,1228+1040 (see Table\,\ref{table:observations} for a full list
of observations). We obtained Infrared Array Camera
\citep[IRAC;][]{fazio04} data in all four channels (3.6 -- 8.0
$\mu$m), Infrared Spectrograph \citep[IRS;][]{houck04} Blue Peak-Up
imaging at 16$\mu$m, and Multiband Imaging Photometer
\citep[MIPS;][]{rieke04} imaging at 24$\mu$m.

The IRAC data reduction was carried out on the Basic Calibrated Data
(BCD) frames as follows. We corrected all of the BCDs for
array-location-dependence using the corresponding correction frames
provided by the Spitzer Science Center (SSC). These were applied after
uncorrecting the array-location-dependence frames for pixel distortion
in order to prevent a repeat correction for pixel distortion when
creating the mosaic. The array-location-corrected BCDs were then all
combined using the standard SSC software, MOPEX \citep[MOsaicker and
  Point source EXtraction]{makovoz06} with dual-outlier rejection to
create a single mosaicked image for each channel. The mosaic pixel
scales were set to the default 1.22\arcsec/pix. As a check on our
reduction method, we also split the BCDs into batches of 7 frames,
creating a mosaic from each batch in order to enable us to test the
repeatability of the photometry and calculate an independent estimate
of our uncertainties.

\begin{figure}
\plotone{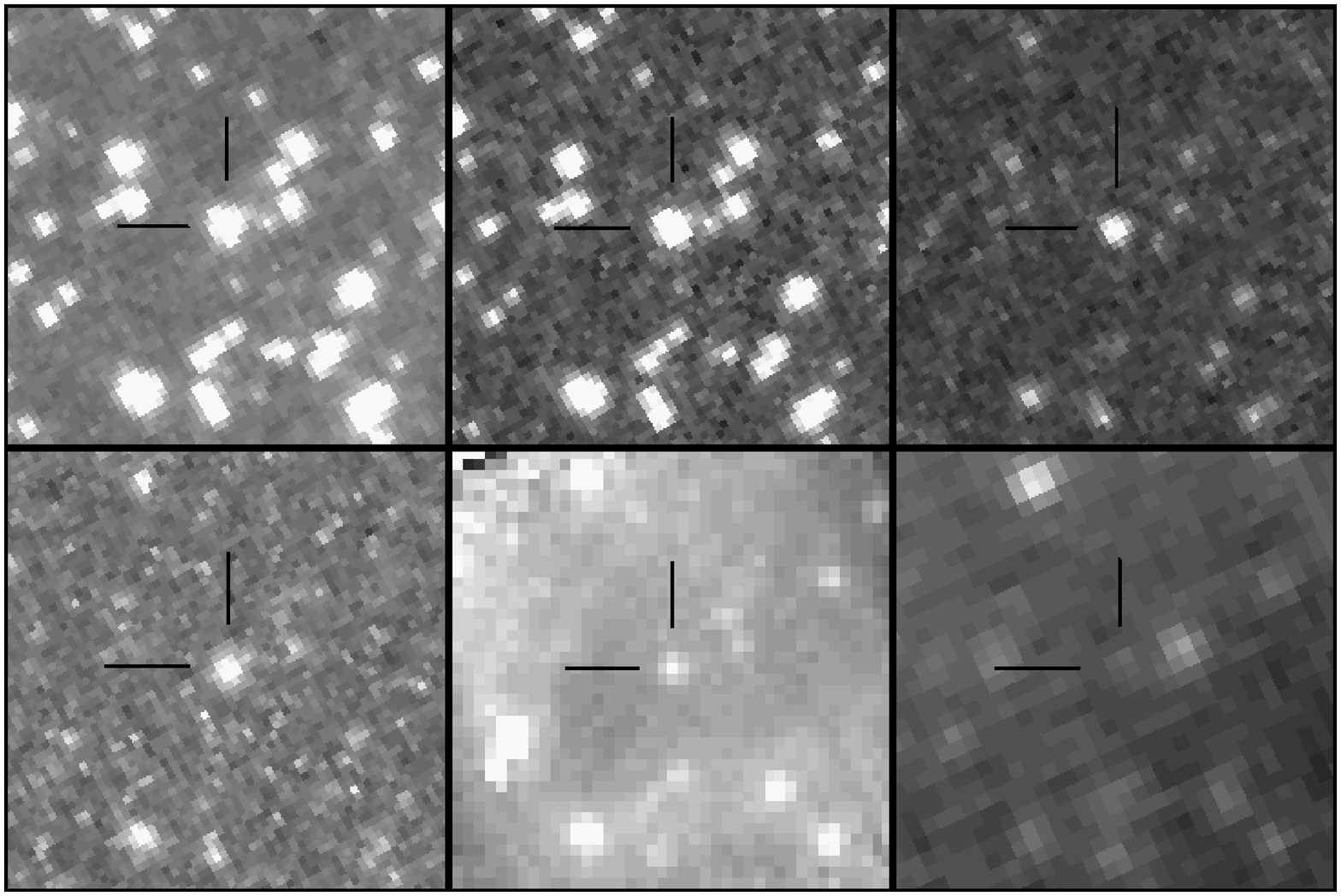}
\caption{Mosaicked images for SDSS\,1228+1040, North up, East left. Top row, left to right: IRAC-1 (3.6$\mu$m), IRAC-2 (4.5$\mu$m), IRAC-3 (5.8$\mu$m). Bottow row, left to right: IRAC-4 (8.0$\mu$m), IRS PU (16$\mu$m), MIPS-24 (24$\mu$m). Pixel scales are 1.22\arcsec/pix, 1.8\arcsec/pix and 2.45\arcsec/pix for the IRAC, IRS PU and MIPS-24 mosaics, respectively. The telescope pointing is good to $<$ 0.5\arcsec}
\label{fig:montage}
\end{figure}

The IRS Peak-Up imaging frames were combined in much the same way,
using MOPEX to combine the BCDs with dual-outlier rejection. Again, we
created a mosaic from both the entire data set, and from sub-sets of
the data, so that we could test the repeatability of the photometry
and derive realistic photometric uncertainties from our data. The
pixel scales for the mosaics were set to the default 1.8\arcsec/pix. A
similar method was applied to the MIPS data, where we created a
separate mosaic for each of the Exposure ID sets (35 BCDs in each
set). This gave us 21 mosaics in total, all of which were created
using multiframe temporal outlier rejection and the default pixel
scale of 2.45\arcsec/pix. The mosaicked images for all of the
\emph{Spitzer} data, with SDSS\,1228+1040 marked, can be found in
Figure\,\ref{fig:montage}.

We carried out aperture photometry on our IRAC and IRS PU mosaics
using IRAF, using apertures of 3.0 pixels for the IRAC mosaics and 2.0
pixels for the IRS mosaics. The sky subtraction was applied using an
annulus of 10--20 pixels in radius. In IRAC channels 1 and 2 there
were background sources included in the sky annuli. IRAF rejects
bright pixels in the sky annuli, but we carried out a double-check of
the sky levels using nearby apertures, and found them to be virtually
identical to the levels originally measured. The photometry was then
converted from MJy/sr to $\mu$Jy and aperture-corrected to the
calibration aperture sizes using the standard aperture corrections
provided by the SSC. The IRAC channel 1 photometry was further
corrected for pixel-phase dependence in accordance with SSC
recommendations. No color correction was applied since we quote the
isophotal effective wavelengths, thus eliminating most of the
color-dependency of the flux calibration and rendering the magnitude
of the correction negligible compared to our uncertainties.

We derived our uncertainty estimates for the IRAC and IRS data using
the standard deviation of the scatter in the sub-set mosaics at each
wavelength. These were then added in quadrature to the absolute
calibration estimates of the IRAC and IRS instruments \citep[][SSC
  website\footnote{http://ssc.spitzer.caltech.edu/irac/documents/reach2005.pdf}]{reach05}.

Looking at the MIPS mosaics, we found that the target is very faint,
and there is bright source within $\sim5-6$ pixels. While it is tempting
to suspect that the bright source may, in fact, be the target, the
pointing of \emph{Spitzer} is good to within 0.5\arcsec, while the
pixel scale of the mosaic is 2.45\arcsec/pix, and so there is no doubt
that the fainter of the two sources is SDSS\,1228+1040. This is
corroborated by matching the positions with respect to the surrounding
stars. We used MOPEX to carry out a source extraction and subtraction
on each mosaic, removing all of the high SNR sources from the field
(including the close, bright neighbour), before carrying out IRAF
aperture photometry on our lower SNR, unsubtracted target. We used an
aperture of 3\arcsec (1.224 pixels) to extract the photometry, with a
background annulus of 20--32\arcsec, and applied the standard aperture
corrections as given in the MIPS Data Handbook. We double-checked the
flux density we obtained by performing MOPEX PRF-fitting on our target
in the original mosaics.  The two methods agreed to within 10\%, while
the standard deviation in the spread in flux densities from the 21
mosaics gave us uncertainties of $\sim$50\%.  The final flux density
was taken as the unweighted median in the 21 measurements from the
aperture photometry.

\subsection{ISAAC data}
The infrared spectroscopic data were taken with the Infrared
Spectrometer And Array Camera (ISAAC) mounted at Antu (UT1), VLT,
Paranal, Chile. The short-wavelength grism was used in its
low-resolution mode at central wavelengths 1.06\,$\mu$m ($z$),
1.25\,$\mu$m ($J$), 1.65\,$\mu$m ($H$), and 2.20\,$\mu$m ($K$). An
$0.8\arcsec$ slit width resulted in resolving powers in the range
650--750. The observations were conducted in service mode on two
nights, 2007 April 03 ($J$ and $H$) and 2007 April 07 ($z$ and $K$).
Flat fields and wavelength calibration with a Xe-Ar lamp (Ar only for
the $z$ spectra) were taken at the start of the night, and telluric
standards were observed within 1 h and an airmass difference $\Delta
M(z) = 0.2$ of the target spectra. The data were obtained as sets of
22 spectra per central wavelength with individual exposure times of
100 s. The individual spectra were taken at different positions on the
detector (AB cycles plus jittering) to minimize the effect of bad
pixels and cosmic rays.

For the reduction, IRAF packages were applied. As first steps, the
data were flatfielded and ``straightened'' via a two-dimensional
wavelength calibration, thus correcting for the positional dependence
of the dispersion.  Corresponding AB pairs were subtracted from each
other, and the resulting data then were combined to a single
image. Subsequently, the target and standard spectra were
extracted. In the case of the $z$ spectrum it proved necessary to
adjust the wavelength calibration using the night sky lines
\citep{rousselot00}.

The following stars were used as telluric standards: HD~106807 (A1V,
$T_\mathrm{eff} = 9\,400$\,K) for the $z$ data, Hip~43699 (B4V,
17\,000\,K) for $J$ and $H$, and Hip~88857 (B3V, 19\,000\,K) for
$K$. The intrinsic lines of the standards were fitted with a Voigt
profile and subtracted from the infrared spectrum, to yield a pure
atmospheric absorption spectrum. For the $z$ data, the intrinsic lines
were submerged in atmospheric absorption bands and could not be
corrected for. The target spectra were then divided by their
corresponding telluric spectra, after the latter had been shifted and
scaled to match the position and depths of the atmospheric features in
the target data. Finally, the spectra were multiplied with a blackbody
corresponding to the effective temperature of the telluric standard,
thus recovering the intrinsic SED.

We used the ISAAC $JHK$ acquisition images to perform differential
photometry against nearby bright stars that have reliable magnitudes
in 2MASS, where we assumed a conservative 10\,\% error for the IR
fluxes derived from these measurements.  The $JHK$ magnitudes of
SDSS\,1228+1040 obtained in this way were then used to correct the
flux level of the ISAAC spectra for slit losses.

\subsection{UKIDSS data}

SDSS\,1228+1040 is within the footprint of the Large Area Survey
within the third Data Release (DR3) of the UKIRT Infrared Deep Sky
Survey \citep[UKIDSS]{lawrence07}.  We obtained $YJHK$ aperture
magnitudes from the WFCAM archive \citep{hambly08}, and converted them
to fluxes at the effective wavelengths using the zero-point
definitions in \citet{hewett06}. The agreement between the UKIDSS and
ISAAC $JHK$ fluxes is very good.

\section{Results and Modelling} \label{sec:models}

\begin{figure}
\plotone{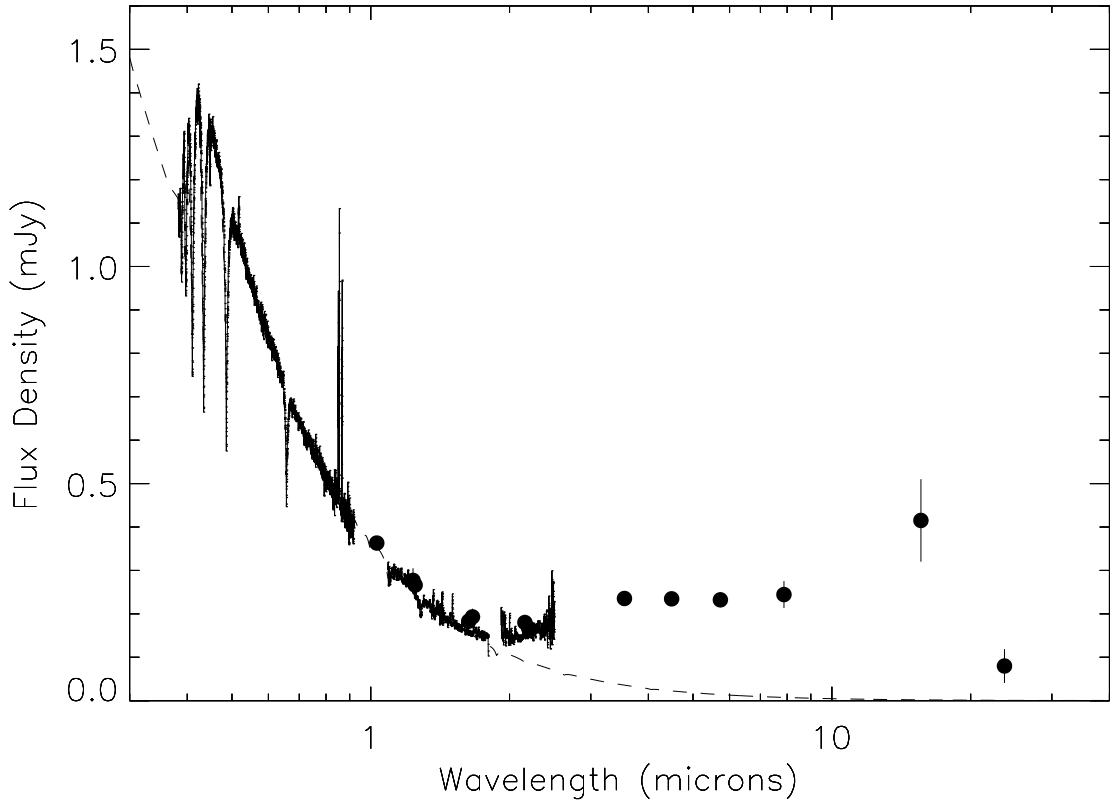}
\caption{Spectral Energy Distribution (SED) for SDSS\,1228+1040 from
  the ultraviolet to the mid-infrared. UV spectra and the white dwarf
  model (dashed line) are taken from \citet{gaensicke06}. The near-IR
  spectroscopy and $YJHK$ photometry were taken with ISAAC/VLT and
  UKIDSS. The mid-IR observations from 3.6 -- 24$\mu$m were taken with
  the \emph{Spitzer Space Telescope} IRAC, IRS Blue Peak-Up and MIPS
  instruments. The SED shows a significant flux density excess from
  the K-band to longer wavelengths, compared to the white dwarf
  alone.}
\label{fig:sdss1228_sed}
\end{figure}

The results of the \emph{Spitzer}, UKIDSS and ISAAC photometry are
given in Table\,\ref{table:photometry}, along with their
uncertainties. Figure\,\ref{fig:sdss1228_sed} shows the \emph{Spitzer}
and ISAAC photometry, and the ISAAC spectroscopy plotted on the same
axes as the UV and optical data from \citet{gaensicke06}. The model of
the expected flux from the white dwarf (taken from the same paper) is
also shown. There is an substantial flux excess over the expected
emission from the white dwarf, starting in the $K$ band and increasing
out to 16$\mu$m, before dropping at 24$\mu$m.

We attempted to model this excess with a simple optically thick,
geometrically thin dust disk model with a temperature profile
$T_\mathrm{disk} \propto r^{-\beta}$, with $\beta=3/4$, as previously
used e.g. by \citet{jura03, becklin05} and \citet{brinkworth07}. In
principle, $\beta$ can take values different from 3/4 depending on the
size of the dust grains, as the grains will cool inefficently once
that the grain size is comparable or smaller than the peak wavelength
of the Planck function at the grain temperature, $r_\mathrm{grain}\la
b/T$ (with $b$ the Wien constant, for details, see
\citealt{brinkworth07}). In practice, however, we found that values
significantly different from 3/4 do not produce viable results. As in
\citet{brinkworth07}, we therefore adopt this value for our
modelling. The input parameters are given in
Table\,\ref{table:modelparams}.

\begin{figure}
\plotone{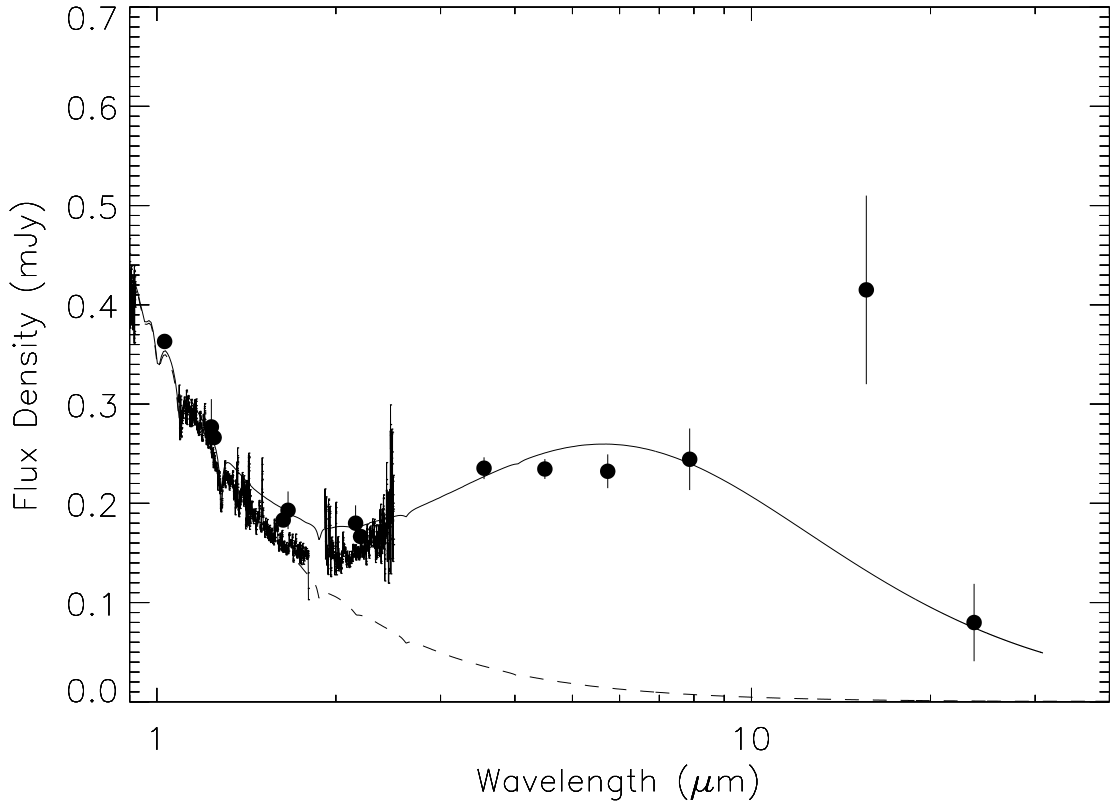}
\caption{The infrared spectral energy distribution of SDSS\,1228+1040, compared with a model consisting of a white dwarf (dashed line) and an optically thick, geometrically thin circumstellar dust disk (dash-dot-dot-dot line). The significant excess at 16$\mu$m may be due to a strong silicate emission feature. For more details see Section\,\ref{sec:models}.}
\label{fig:optthick}
\end{figure}

The model is shown in Figure\,\ref{fig:optthick}. While it fits the
near-IR, IRAC and MIPS points well, the 16 micron flux density is 2.0
sigma greater than predicted. This 16 $\mu$m excess may be explained
by the presence of a significant silicate emision feature at 10$\mu$m,
as seen in the similar dusty white dwarf GD\,362. \citet{jura07a}
found that this emission feature has a strong red wing that, if
present in SDSS\,1228+1040, would contribute significantly to the IRS
Peak-Up band (which has a passband from 13 -- 18.5 $\mu$m). It is
possible that such a feature would also affect the 8 $\mu$m flux
density, but the blue wing of the silicate feature in GD\,362 was less
pronounced than the red wing, so the effect is expected to be minor
(of order a few \%). This is not currently accounted for in the model,
but could explain the slight rise seen at 8 $\mu$m from the rest of
the IRAC bands.


\section{Discussion and Conclusions}

The ISAAC, UKIDSS and \emph{Spitzer Space Telescope} data show a
noticeable infrared excess in SDSS\,1228+1040, in excess of the flux
density expected from the white dwarf alone. The observed spectral
energy distribution provides clear and compelling evidence for the
presence of a substantial dusty component to the debris disk, in addition to the
gaseous component discovered by \citet{gaensicke06}, indicating that
both gaseous and dusty debris material can coexist around a white
dwarf. The presence of dust around SDSS\,1228+1040 implies that debris
disks can survive around relatively hot parent white dwarfs, contrary
to the earlier suggestion by \citet{kilic06, vonhippel07}, based on a
smaller sample of white dwarfs with debris disks, that white dwarfs
hotter than around 15,000\,K should sublimate any orbiting dust disk.

Our model shows that
an optically thick dust disk model can reproduce the SED seen in
SDSS\,1228+1040, except for the 16 $\mu$m point, which we currently
attribute to the red wing of a strong silicate feature at 10$\mu$m, as
seen in similar systems (e.g. GD 362). We calculated a lower-limit estimate of the 
mass of dust in the optically thick disk using the generic mass absorption 
coefficient used by \citet{beckwith90} of:

\[\kappa_{\nu} = 0.1(\nu/10^{12} Hz)^\beta \rm{cm^2 g^{-1}}, 
\]

where we follow their example and adopt $\beta = 1$ \citep[see][for a more in-depth discussion]{brinkworth07}.
 This leads to a disk mass of at least 1 $\times 10^{22}$ g, equivalent to $\sim$1.3 $\times 10^{-4} M_{Moon}$ or $\sim$0.01 times the mass of Ceres. 

There are, however, two main issues with our disk model. Firstly, in order
to match the rise from the $K$ band to the IRAC channel 1 point at
3.6$\mu$m, we find that the inner temperature of the dust disk has to
be around 1670\,K, which is hotter than the expected sublimation
temperature of the dust grains (around 1200--1400\,K). Moreover, if we
estimate the inner radius of the gaseous disk from \citet{gaensicke06}
using $R_\mathrm{inner~disk}\sim (V_\mathrm{peak}/V_\mathrm{max})^2 \times
R_\mathrm{outer~disk}$,
where $V_\mathrm{peak}$ is the velocity of the line peaks and
$V_\mathrm{max}$ is the maximum velocity at which emission is seen,
then we find that the inner radius of the dusty disk at this
temperature is smaller (18\,R$_\mathrm{WD}$) than the inner radius of the
gaseous disk ($\sim$27\,R$_\mathrm{WD}$). It is possible that the inclusion of
a 2- or 3-zone, or warped disk model would allow us to fit the data
using a cooler inner disk temperature (and therefore larger inner dust
disk radius) but the lack of data points in the mid-IR would make it
impossible to usefully constrain such a model, and so we do not pursue
it further.

Secondly, we find that the implied outer radius of the dusty disk
($\sim 107\,R_\mathrm{WD}$) coincides almost perfectly with the outer
radius of the optically thin gaseous disk ($\sim
108$\,$R_\mathrm{WD}$), as derived by \citet{gaensicke06}, so the
gaseous dust disk permeates throughout the dusty component. This is
not unexpected in itself, as the sublimating dust will likely
replenish the gas disk. It is possible, however, that the lifetime of the
dust grains will be relatively short due to the increased viscosity. \citet{gaensicke07} determined the photospheric abundance of Mg in SDSS1228+1040 to be $\log(\mathrm{Mg/H})=-4.58\pm0.06$. Using the prescription of \citet{koester06}, and assuming that hydrogen fraction in the circumstellar material is depleted by a factor one hundred with respect to its solar value, we estimate an accretion rate of $\dot M\simeq3\times10^{9}\mathrm{g\,s^{-1}}$. This is within the range of accretion rates obtained by \citet{farihi08b} following a similar approach for the currently known sample of metal-polluted white dwarfs with dusty debris disks. We stress, however, that this estimate of the accretion rate should is likely to be a strict upper limit, as $\dot M$ is inverse proportional to the hydrogen depletion in the circumstellar material, and the factor 100 assumed above is probably underestimated \citep{friedrich99,gaensicke08}. Combining the accretion rate with the estimate of the disk mass from above, we estimate a lower limit on the life time of the disk of $\sim1000$\,yr.

\vspace{20mm}

\acknowledgments

We thank the participants of the Winter AAS 2008 White Dwarf Splinter
Session for their extremely helpful discussions and suggestions. This work is based on observations made
with the \emph{Spitzer Space Telescope}, which is operated by the Jet
Propulsion Laboratory, Caltech, under NASA contracts 1407 and 960785.
This work makes use of data products from the Two Micron All Sky
Survey, which is a joint project of the University of Massachusetts
and IPAC/Caltech, funded by NASA and the NSF. We acknowledge and thank the
referee Jay Farihi for his detailed report.

\clearpage

\clearpage
\begin{deluxetable}{lccccc}
\tablecaption{Observation log for Spizer Space Telescope data}
\tablehead{
\colhead{Instrument/} & \colhead{Wavelength} & \colhead{AOR Key} & \colhead{Total Integration Time} & \colhead{Date} & \colhead{Pipeline} \\
\colhead{Channel} & \colhead{microns} & \colhead{} & \colhead{(s)} & \colhead{} & \colhead{} \\
}
\tablecolumns{6}
\startdata
IRAC Ch. 1 & 5.550 & 22247936 & 1000 & 2007-06-30 & S16.1 \\
IRAC Ch. 2 & 4.493 & 22247936 & 1000 & 2007-06-30 & S16.1 \\
IRAC Ch. 3 & 5.731 & 22247936 & 1000 & 2007-06-30 & S16.1 \\
IRAC Ch. 4 & 7.872 & 22247936 & 1000 & 2007-06-30 & S16.1 \\
IRS Blue Peak-Up & 15.6 & 22248192 & 3450 & 2007-06-16 & S16.1 \\
MIPS-24    & 23.68 & 25459200  & 3012 & 2008-07-25 & S18.1 \\
MIPS-24    & 23.68 & 25459456  & 3012 & 2008-07-25 & S18.1 \\
MIPS-24    & 23.68 & 25459712  & 3012 & 2008-07-25 & S18.1 \\
\enddata \label{table:observations}
\end{deluxetable}
\clearpage

\clearpage
\begin{deluxetable}{lccc}
\tablecaption{Photometry for SDSS\,1228+1040 from \emph{Spitzer} and ISAAC}
\tablehead{
\colhead{Instrument/} & \colhead{Wavelength} & \colhead{Flux Density} & \colhead{Uncertainty} \\ 
\colhead{Band}           & \colhead{$\mu$m}      & \colhead{mJy}          & \colhead{\%} \\ 
}
\tablecolumns{4}
\startdata
ISAAC / J & 1.235 & 0.277 & 10 \\
ISAAC / H & 1.662 & 0.193 & 10 \\
ISAAC / K & 2.159 & 0.180 & 10 \\
UKIDSS / y & 1.031 & 0.363 & 1.1 \\
UKIDSS / J & 1.248 & 0.266 & 1.5 \\
UKIDSS / H & 1.631 & 0.183 & 2.2 \\
UKIDSS / K & 2.101 & 0.167 & 3.6 \\
IRAC / 1  & 3.550 & 0.235 & 4.5 \\
IRAC / 2  & 4.493 & 0.235 & 4.1 \\
IRAC / 3  & 5.731 & 0.232 & 7.5 \\
IRAC / 4  & 7.872 & 0.244 & 12.8 \\
IRS PU / 16  & 15.6  & 0.415 & 23.0 \\
MIPS / 24 & 23.68 & 0.080 & 49 \\

\enddata \label{table:photometry}
\end{deluxetable}
\clearpage

\clearpage
\begin{deluxetable}{ccccc}
\tablecaption{Parameters for the dust disk model discussed in Section\,\ref{sec:models}.}
\tablehead{}
\tablecolumns{5}
\startdata
Inner Disk Temperature (K) & 1670 \\
Inner Disk Radius (R$_\mathrm{WD}$) & 18 \\
Outer Disk Temperature (K) &  450  \\
Outer Disk Radius (R$_\mathrm{WD}$) & 107  \\
Disk Inclination (degrees) & 70 \\

\enddata \label{table:modelparams}
\end{deluxetable}
\clearpage


\begin{thebibliography}{99}

\bibitem[Adelman-McCarthy et al.(2006)]{adelman-mccarthy06} Adelman-McCarthy, J.\,K., Ag\"ueros, M.\,A., Allam, S.\,S., Anderson, K.\,S.\,J., Anderson, S.\,F., Annis, J., Bahcall, N.\,A., Baldry, I.\,K., Barentine, J.\,C., Berlind, A. and 131 coauthors, 2006, ApJS, 162, 38

\bibitem[Becklin et al.(2005)]{becklin05} Becklin, E.~E., Farihi, J., Jura, M., Song, I., Weinberger, A.~J., \& Zuckerman, B.\ 2005, \apjl, 632, L119

\bibitem[Beckwith et al.(1990)]{beckwith90} Beckwith, S.\,V.\,W., Sargent, A.\,I., Chini, R.\,S. \& G\"usten, R. 1990, AJ, 99, 924

\bibitem[Brinkworth et al.(2007)]{brinkworth07} Brinkworth, C.\,S., Hoard, D.\.W., Wachter, S., Howell, S.\,B., Ciardi, D.\,R., Szkody, P., Harrison, T.\,E., van Belle, G.\,T., Esin, A.\,A. 2007, ApJ, 659, 1541

\bibitem[Debes \& Sigurdsson(2002)]{debes02} Debes, J.\,H., Sigurdsson, S. 2002, ApJ, 572, 556

\bibitem[Fazio et al.(2004)]{fazio04} Fazio, G., et al.\ 2004, \apjs, 154, 10

\bibitem[Farihi et al.(2008a)]{farihi08a} Farihi, J., Zuckerman, B., Becklin, E.\,E., 2008, ApJ, 674, 431 

\bibitem[Farihi et al.(2008b)]{farihi08b} Farihi, J., Jura, M., Zuckerman, B. \ 2008, ApJ in press (arXiv:0901.0973)

\bibitem[Friedrich et al.(1999)]{friedrich99} Friedrich, S., Koester, D., Heber, U., Jeffery, C. S., Reimers, D. \ 1999, A\&A, 350, 865

\bibitem[G\"ansicke et al.(2006)]{gaensicke06} G\"ansicke, B.\,T., Marsh, T.\,R., Southworth, J., Rebassa-Mansergas, A. \ 2006, Science, 314, 1908

\bibitem[G\"ansicke et al.(2007)]{gaensicke07} G\"ansicke, B.\,T., Marsh, T.\,R., Southworth, J. \ 2007, MNRAS, 380, L35

\bibitem[G\"ansicke et al.(2008)]{gaensicke08} G\"ansicke, B.\,T., Koester, D., Marsh, T.\,R., Rebassa-Mansergas, A., Southworth, J. \ 2008, MNRAS, 391, L103 

\bibitem[Graham et al.(1990)]{graham90} Graham, J.\,R., Matthews, K., Neugebauer, G., Soifer, B.\,T., 1990, ApJ, 357, 216

\bibitem[Hambly et al.(2008)]{hambly08} Hambly, N.\,C., Collins, R.\,S., Cross, N.\,J.\,G., Mann, R.\,G., Read, M.\,A., Sutorius, E.\,T.\,W., Bond, I., Bryant, J., Emerson, J.\,P., Lawrence, A. and 7 co-authors. 008, MNRAS, 384, 637

\bibitem[Hewett et al.(2006)]{hewett06} Hewett, R.\,J., Warren, S.\,J., Leggett, S.\,K., Hodgkin, S.\,T. 2006, MNRAS, 367, 454

\bibitem[Hoard et al.(2007)]{hoard07} Hoard, D.\,W., Howell, S.\,B., Brinkworth, C.\,S., Ciardi, D.\,R., Wachter, S. 2007, ApJ, 671, 734

\bibitem[Houck et al.(2004)]{houck04} Houck, J. et al. 2004, ApJSS, 154, 18

\bibitem[Jura(2003)]{jura03} Jura, M. \ 2003, ApJ, 584, L91

\bibitem[Jura et al.(2007a)]{jura07a} Jura, M., Farihi, J., Zuckerman, B. Becklin, E. E. 2007, AJ, 133, 1927

\bibitem[Jura et al.(2007b)]{jura07} Jura, M., Farihi, J., Zuckerman, B. \ 2007, ApJ, 663, 1285

\bibitem[Jura et al.(2008)]{jura08} Jura, M., Farihi, J., Zuckerman, B. \ 2008, AJ in press (arXiv:0811.1740)

\bibitem[Jura(2008)]{jura08a} Jura, M., 2008, AJ, 135, 1785

\bibitem[Kilic et al.(2005)]{kilic05} Kilic, M., von Hippel, T., Leggett, S.\,K., Winget, D.\,E. \ 2005, ApJ, 632, L115

\bibitem[Kilic et al.(2006)]{kilic06} Kilic, M., von Hippel, T., Leggett, S.\,K., Winget, D.\,E. \ 2006, ApJ, 646, 474

\bibitem[Koester et al.(1997)]{koester97} Koester, D., Provencal, J., Shipman, H.\,L. \ 1997, A\&A, 320, L57

\bibitem[Koester \& Wilken(2006)]{koester06} Koester, D., Wilken, D. 2006, A\&A, 453, 1051

\bibitem[Makovoz et al.(2006)]{makovoz06} Makovoz, D., Roby, T., Khan, I., Booth, H., 2006, SPIE, 6274, 10

\bibitem[Lawrence et al.(2007)]{lawrence07} Lawrence, A., Warren, S.\,J., Almaini, O., Edge, A.\,C., Hambly, N.\,C, Jameson, R.\,F., Lucas, P., Casili, M., Adamson, A., Dye, S. and 12 co-authors. 2007, MNRAS, 379, 1599

\bibitem[Reach et al.(2005)]{reach05} Reach, W.\,T., Kuchner, M.\,J., von Hippel, T., Burrows, A., Mullally, F., Kilic, M., Winget, D.\,E. \ 2005, ApJ, 635, L161

\bibitem[Rieke et al.(2004)]{rieke04} Rieke, G. et al. 2004, ApJSS, 154, 25

\bibitem[Rousselot et al.(2000)]{rousselot00} Rousselot, P., Lidman, C., Cuby, J.-G., Moreels, G., Monnet, G. 2000, A\&A, 354, 1134


\bibitem[Telesco, Joy \& Sisk]{telesco90} Telesco, C.\,M., Joy, M., Sisk, C. \ 1990, ApJ, 358, L17

\bibitem[Tokunaga et al.(1990)]{tokunaga90} Tokunaga, A.~T., 
Becklin, E.~E., \& Zuckerman, B.\ 1990, \apjl, 358, L21 

\bibitem[von Hippel et al.(2007)]{vonhippel07}von Hippel, T., Kuchner, M.\,J., Kilic, M., Mullally, F., Reach, W.\,T. \ 2007, ApJ, 662, 544

\bibitem[Zuckerman \& Becklin(1987)]{zuckerman87} Zuckerman, B., \& Becklin, E.\,E. \ 1987, Nat., 330, 138



\end{thebibliography}
\end{document}